\documentclass[preprint,12pt]{elsarticle}

\usepackage{amssymb}

\journal{Nuclear Instruments and Methods A}

\begin{document}

\begin{frontmatter}



\title{Characterization of a Nd-loaded organic liquid scintillator for neutrinoless double beta decay search of $^{150}$Nd with a 10-ton scale detector}

\author[inr]{I.~Barabanov}

\author[inr]{L.~Bezrukov}

\author[mib]{C.~Cattadori\corref{cor1}}

\author[ipc]{N.~Danilov}

\author[lngs]{A.~Di Vacri}

\author[lngs]{A. ~Ianni\corref{cor1}}

\ead{carla.cattadori@lngs.infn.it, aldo.ianni@lngs.infn.it}

\cortext[cor1]{Corresponding author}

\author[lngs]{S.~Nisi}

\author[pg]{F.~Ortica}

\author[pg]{A.~Romani}

\author[lngs]{C.~Salvo}

\address[lngs]{INFN LNGS S.S. 17bis, km 18+910, 67010 Assergi, Italy}

\address[mib]{INFN Milano Bicocca, Italy}

\author[jinr]{O.~Smirnov}
\address[jinr]{Joint Inst Nucl Res, Dubna 141980, Moscow Region Russia}

\author[inr]{E.~Yanovich}
\address[inr]{INR-RAS Moscow, Russia}

\address[ipc]{IPC-RAS, Moscow, Russia}

\address[pg]{Dipartimento di Chimica, Universit\`a di Perugia and INFN, 06123 Perugia, Italy}

\begin{abstract}
Several liters of an organic liquid scintillator (LS) loaded with Nd have been made. We report on performances of this scintillator in terms of optical properties, radiopurity and light yield for a Nd concentration of 6.5~g/l. A possible application to search for the $^{150}$Nd neutrinoless double beta ($0\nu\beta\beta$) decay with a 10-ton scale LS detector is discussed together with further improvements.
\end{abstract}


\end{frontmatter}


\section{Introduction}\label{Intro}
In this paper we report on a number of measurements performed to characterize the properties of an organic liquid scintillator based on pseudocumene and loaded with Nd. This work has been carried out within the MetaLS project \cite{MetaLS}. In particular, we will report on optical measurements made in order to determine the scintillator response and on radiopurity measurements and purification methods, to discuss possible applications in the double beta decay searches field. 

Several liters of Nd-loaded organic liquid scintillator (NdLS) have been produced following the methodic developed to produce Yb, In and Gd loaded scintillators \cite{YbIGdLS},\cite{InLS1},\cite{InLS2},\cite{GdLS1},\cite{GdLS2},\cite{GdLS3}. Measurements were carried out in small samples and in a 2.5~l, 1 m long rectangular quartz cell equipped with two 3-inch photomultipliers. Radiopurity measurements were performed by Inductively Coupled Plasma Mass Spectrometry (ICP-MS) and $\gamma$-spectroscopy by High Purity Germanium detector (HP-Ge) facilities at the Gran Sasso Laboratory.

We have also studied and report here on the performances of a 10-ton scale liquid scintillator detector loaded with Nd to search for the  $^{150}$Nd $0\nu\beta\beta$ decay. 

\section{The Nd-loaded organic liquid scintillator and its optical properties}
A 2.5 l batch of NdLS based on Nd carboxilate (Nd-CBX) having a Nd concentration ([Nd]) of 6.5 g/l has been prepared in spring 2008 starting from a concentrated master solution ([Nd] $\sim$ 50 g/l) prepared in December 2007. The adopted Nd concentration is the result of a trade-off between the Nd target mass, the light yield and the attenuation length of the doped LS.  The choice of the carboxylic acid, the proper reagent purification procedures as well as the Nd-CBX solid salt synthesis and drying recipe, are the results of a R\&D activity carried on in the years 2006-2007 and are a follow-up of the know-how developed in our working group to produce large quantities of stable and highly performing metal (Gd and In) loaded liquid scintillators. The ability in producing highly Nd concentrated master solution allows to easily produce NdLS batches of several tens of liters in a chemical research laboratory environment.   

\begin{figure}[h]
\begin{minipage}{14pc}
\includegraphics[width=18pc]{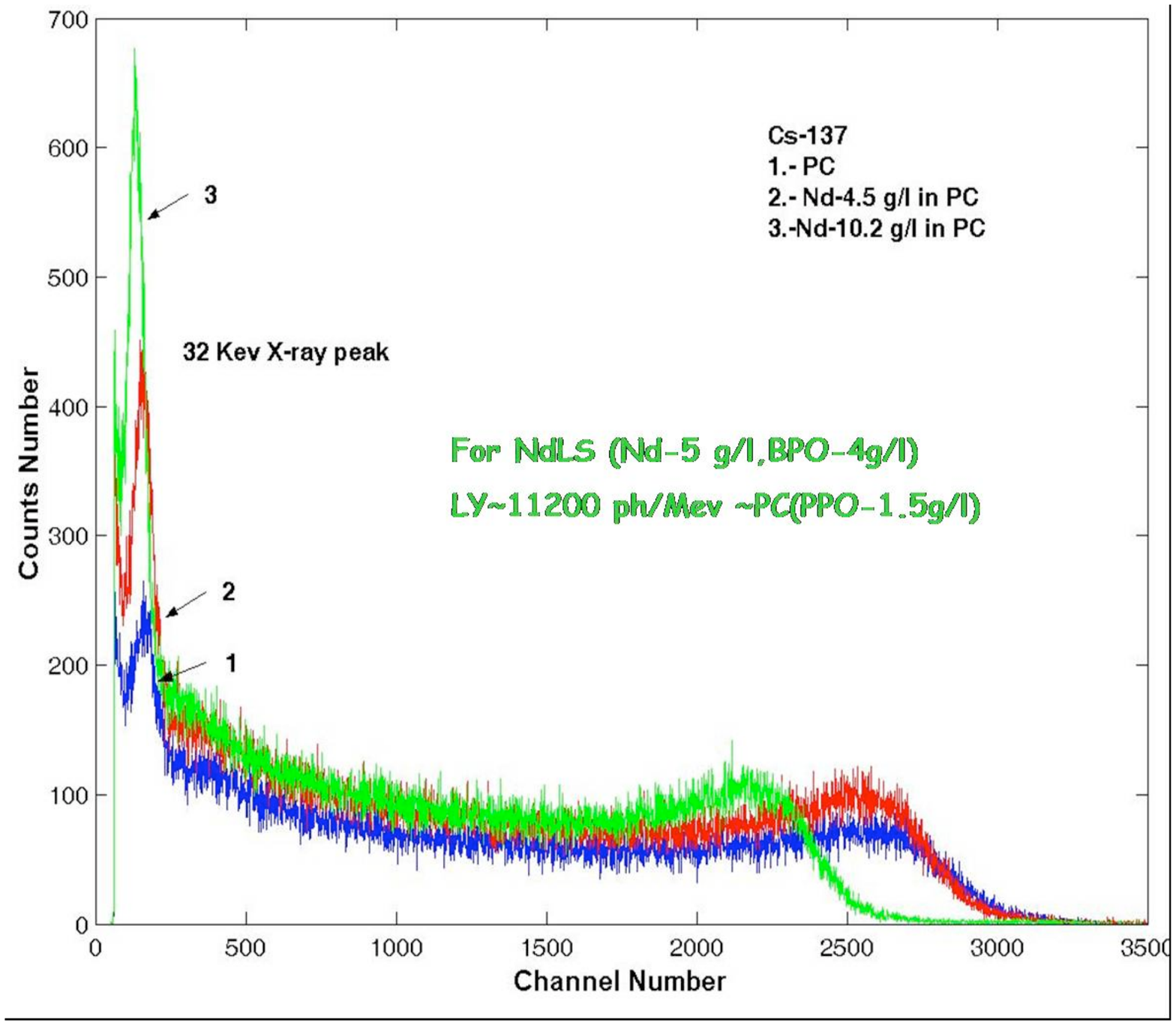}
\caption{\label{LY small vial}LY of NdLS for different Nd concentration of PPO and of the reference LS as measured in small vials. For LS 1.5 g/l of PPO as primary fluor is adopted, while for NdLS BPO at 4.0 g/l. The use of BPO at the adopted concentration leads to a LY larger of 10-20$\%$ than with PPO at 1.5 g/l.}
\end{minipage}\hspace{4pc}%
\begin{minipage}{18pc}
\includegraphics[width=18pc]{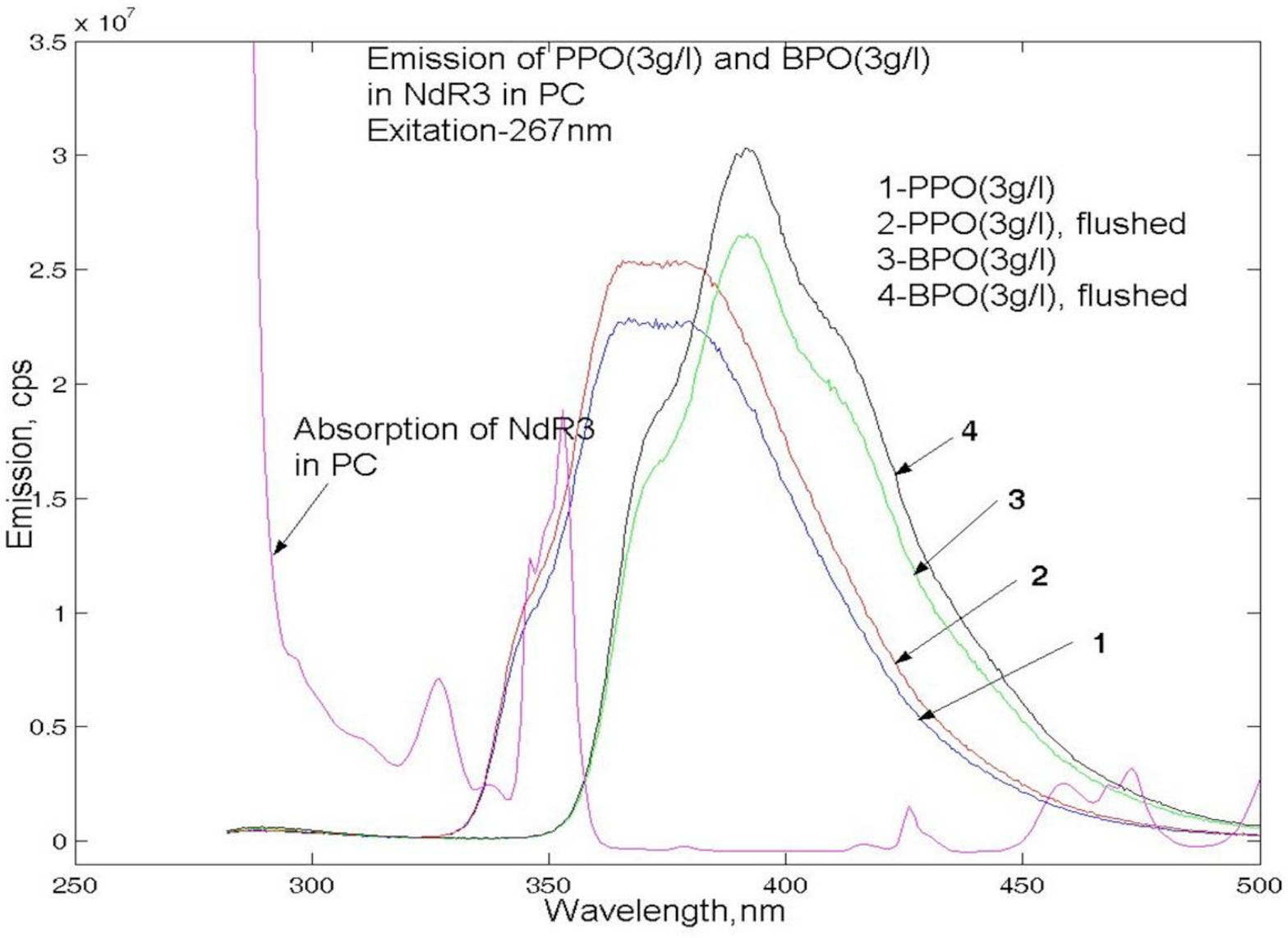}
\caption{\label{assorbanza e emissione}Emission of NdLS with PPO(3g/l) and BPO(3g/l); excitation 267 nm. Nd-CBX in PC absorbance spectrum is superimposed. }
\end{minipage}
\end{figure}

The light Yield (LY) of the prepared NdLS and of the reference LS (pseudocumene based), both using PPO as primary fluor at concentration of 1.5 g/l, was measured in small cylindrical vials (5 ml), optically coupled  to a PMT window by silicon grease. The vials were wrapped by teflon tape to improve the light collection. As reference the LY of PC with the same fluor concentration, measured in the same vial, is adopted ($\sim$ 10000 ph/MeV). The vial is irradiated by a $^{137}$Cs source  (E$_{\gamma}$ = 662 keV). The measured LY values come from the comparison of the NdLS to the reference LS Compton spectra as shown in Fig.\ref{LY small vial}.  The measured light yield, at [Nd]=6.5~g/l and [PPO]=1.5 g/l, is $\sim$~75\% of that of PC based LS at the same fluor concentration. 

The optical characterization of the 2.5 l NdLS batch is completed by a spectrophotometric measurement of the light absorption spectra in the relevant wavelength range (300--450 nm) and by fluorimetric measurement of the light emission spectra. A challenge is the presence of Nd characteristic absorption bands, and of other colored rare earths that usually come with Nd, in the wavelength region where PMTs quantum efficiency is maximal; this cause scintillation light loss. For this reason, particular attention has been put on the purification from optically active and from radioactive contaminants in view of the specific scientific application (0$\nu\beta\beta$ decay search). Adopting our reagent purification procedure, we are able to obtain a transparent region around 400~nm as shown in Fig.\ref{assorbanza e emissione}. In the same figure, the emission spectra of NdLS prepared with the same concentrations of either PPO or BPO as primary fluors, are superimposed; the emission spectra match quite well the NdLS transparent region.

\section{Light Yield and attenuation length measurement}

\begin{figure}[t]
\centerline{
\includegraphics[width=0.6\textwidth,angle=0]{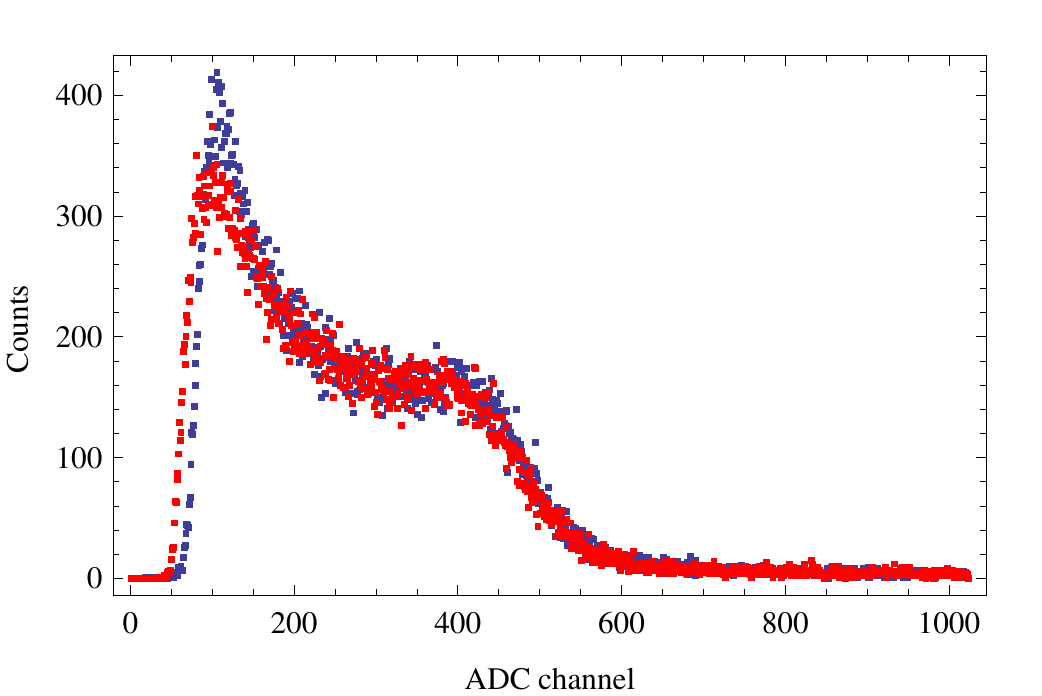}
}
\caption{
Charge spectra of both PMTs in the 1-m cell test when the $^{137}$Cs source is loacated at the center. 
\label{spectra_at_center}}
\end{figure}

Besides the measurements described above we have also performed, starting in spring 2008, tests with a 1-m long rectangular quartz cell filled with scintillator. The cell has been filled first with undoped PC-based (LS) scintillator (PC, [PPO]=1.5 g/l) as reference, then with NdLS (Nd-CBX in PC, [Nd]=6.5 g/l, [PPO]=1.5 g/l,[MSB]= 30 mg/l).  The goal is to measure on a few liters scale: 
\begin{itemize}
 \item the light yield to be compared to the value obtained from several milliliters samples
 \item to measure the attenuation length which is a fundamental parameter for any application of the NdLS. 
\end{itemize}
The cell is equipped with two 3-inch photomultipliers (PMTs) optically coupled to the quartz by silicon grease at the two end sides. The cell is wrapped with a VM2000 sheet to increase the light collection. The two PMTs are put in coincidence and their charge values are stored. The cell is irradiated, scanning it along its main axis, by a collimated 100~kBq $^{137}$Cs  source. For each source position, we build up the Compton spectrum for each PMT; the half-height of the Compton shoulder, $L_{1/2}^{CE}$,  is used to compare data with and without the Nd.  The scintillator light attenuation length is then estimated, plotting the quantity $L_{1/2}^{CE}$ against the relative distance to one PMT and fitting with the function: 
\begin{equation}
y(x) = \frac{k_1}{x^2} + k_2e^{-\frac{x}{\lambda}}  \label{funzione fit}
\end{equation}
The first term reproduces the direct light collection, relevant close to the PMTs, while the second the light collection attenuated by the NdLS. $k_{1,2}$ and $\lambda$ are unknowns and determined from the data. In Fig.~\ref{spectra_at_center} we show the charge spectra of the two PMTs when the source is located at the center of the cell. 

\begin{figure}[h]
\begin{minipage}{14pc}
\includegraphics[width=18pc]{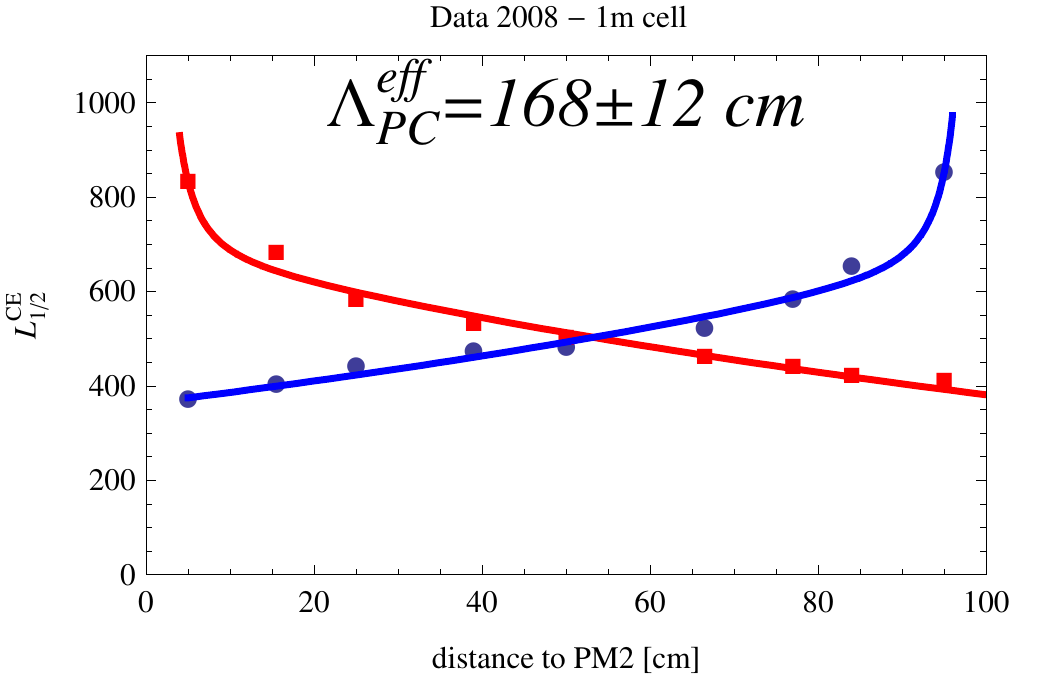}
\caption{\label{PC2008}Distribution of $L_{1/2}^{CE}$ for both PMTs against the relative distance to PMT2 (see text for detailes) for LS wihtout Nd. 2008 data.}
\end{minipage}\hspace{4pc}%
\begin{minipage}{18pc}
\includegraphics[width=18pc]{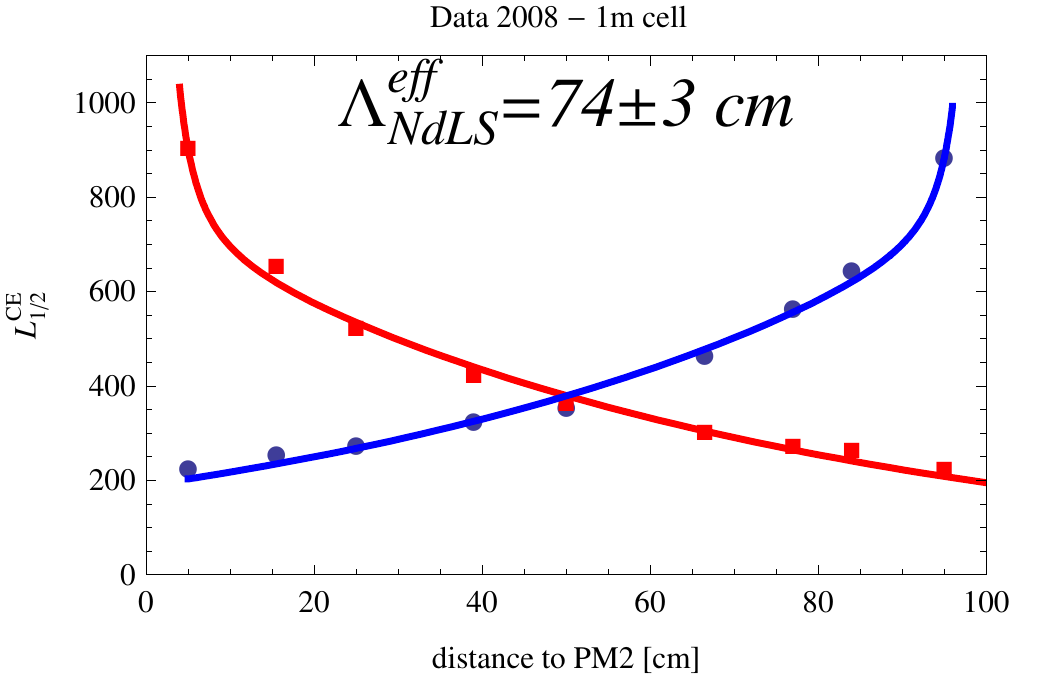}
\caption{\label{Nd2008}Distribution of $L_{1/2}^{CE}$ for both PMTs against the relative distance to PMT2 (see text for detailes) for LS wiht Nd. 2008 data. }
\end{minipage}
\end{figure}

In Fig.~\ref{PC2008} and \ref{Nd2008} we show the results of measurements performed in 2008 with and without Nd, respectively. When fitting the data sets with the function (\ref{funzione fit}), we determine:
\begin{description} 
\item $\Lambda^{eff}_{PC}=168\pm12$ 
\item $\Lambda^{eff}_{NdLS}=74\pm3$
\end{description}

\begin{figure}[h]
\begin{minipage}{14pc}
\includegraphics[width=18pc]{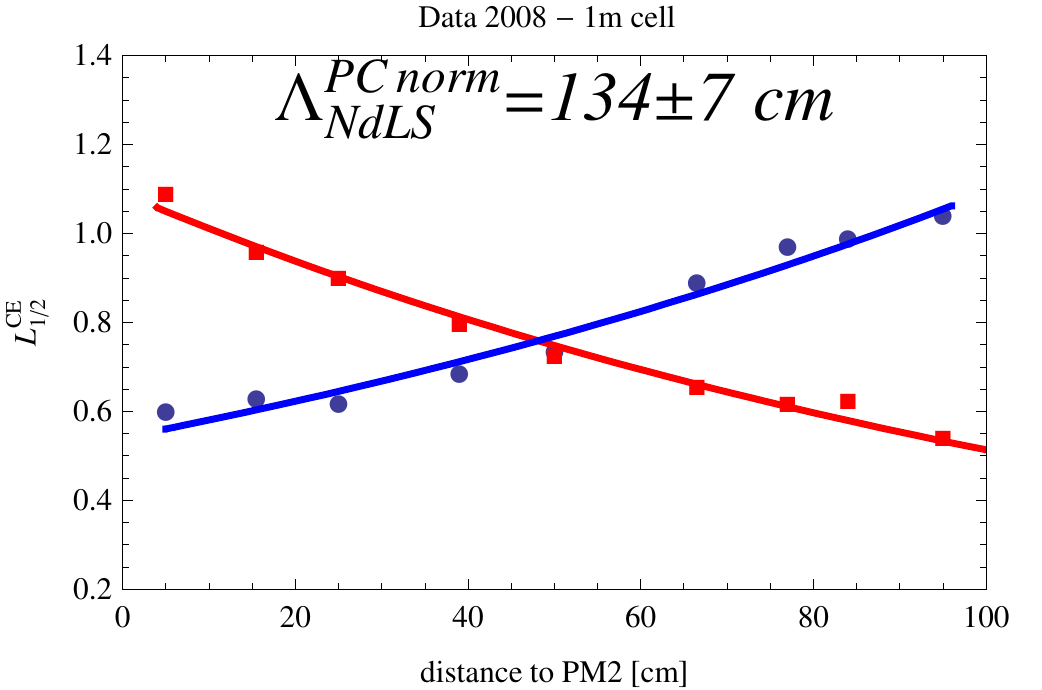}
\caption{\label{FitPCnorm}Distribution of $L_{1/2}^{CE}$ for measurement taken with the NdLS and normalizing data to corresponding PC measurements (see text for details).}
\end{minipage}\hspace{4pc}%
\begin{minipage}{18pc}
\includegraphics[width=18pc]{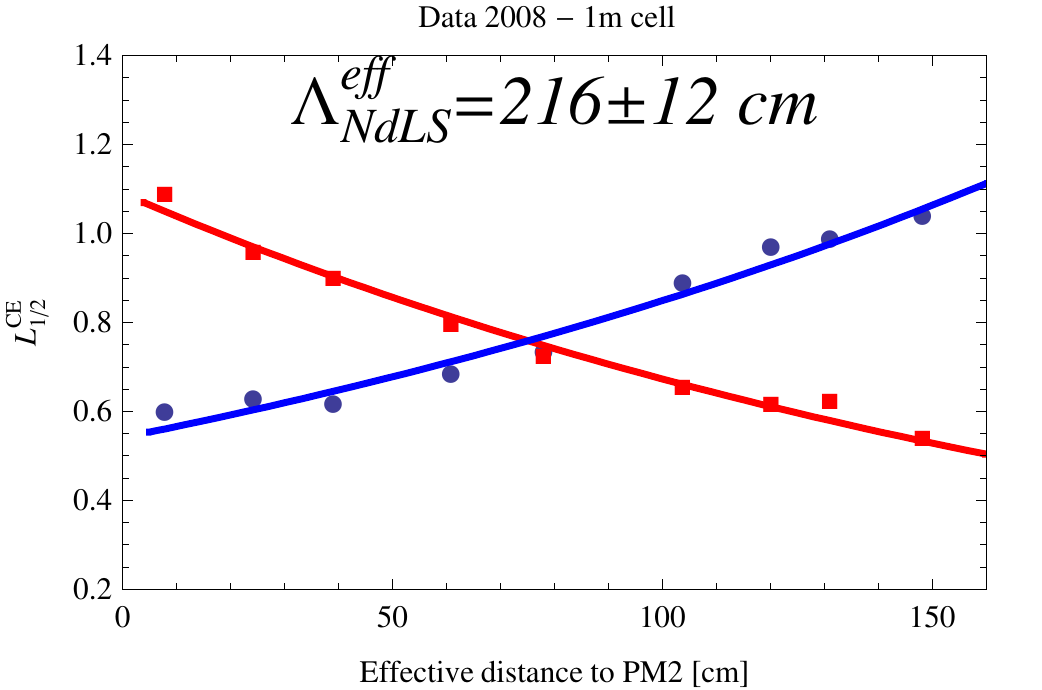}
\caption{\label{FitNdeff}Distribution of $L_{1/2}^{CE}$ for measurements taken with the NdLS and using the actual cell optical path from timing data.}
\end{minipage}
\end{figure}

As the Borexino grade PC LS attenuation length is known to be of the order of $\sim$8m~\cite{BorexinoNIM440}, and we measure it to be $\sim$8m when measuring by spectrophotometry in 10cm long cell prior the 1m cell PC filling, it is evident that the value obtained from the  1m cell measurements is affected by geometric effects related to light propagation. Therefore to extract the NdLS light attenuation length from our data we adopt the following procedure:
\begin{enumerate}
\item we normalize the NdLS $L_{1/2}^{CE}$ to the PC $L_{1/2}^{CE}$ and we plot it as a function of the PMT to source distance. It is shown in Fig.~\ref{FitPCnorm}.  The exponential decay trend can now be attributed to the light absorbance of the NdLS. 
\item we determine the effective cell optical path taking into account the light reflection at the external quartz surface: after equalizing in timing and in amplitude the PMT signals when the source is placed at the cell central position, we then move the source at one end. From PMT signal timing measurements we can evaluate the effective optical path which turns out to be 1.56 m. 
\item We correct the geometrical PMT to source distance for the actual optical path. 
\end{enumerate}
When doing this, the data scale as shown in Fig.~\ref{FitNdeff}. The fitting of the rescaled data set leads to an attenuation length of $216\pm12$~cm for our NdLS at [Nd]= 6.5~g/l.\\
A test with a longer cell and a full optical modeling of the light propagation is planned.

\begin{figure}[t]
\centerline{
\includegraphics[width=0.6\textwidth,angle=0]{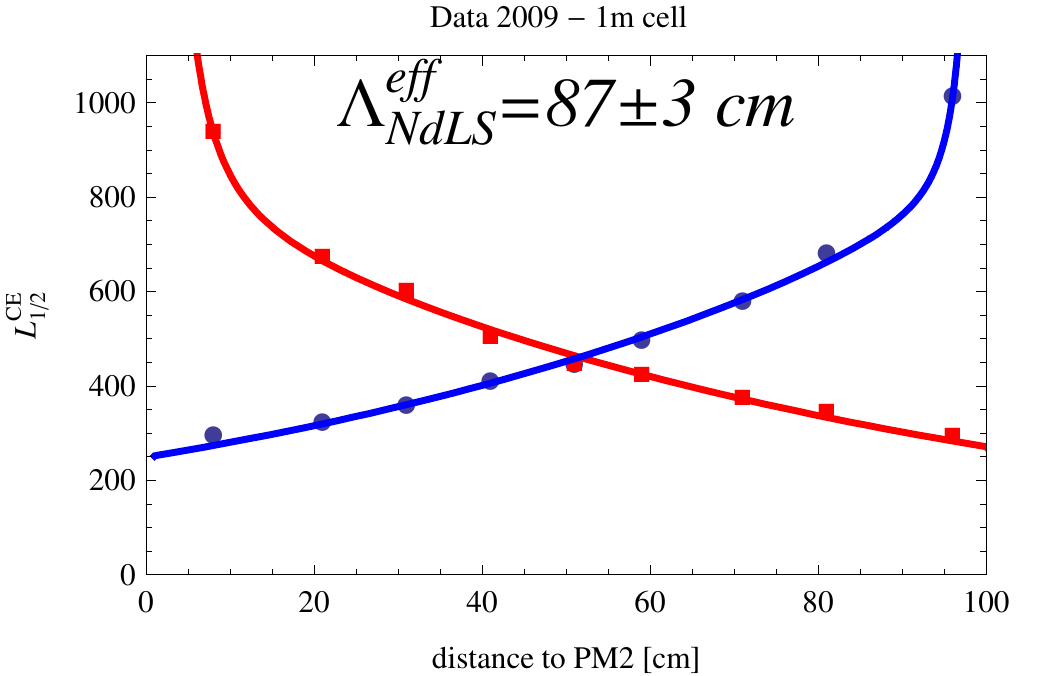}
}
\caption{
Distribution of $L_{1/2}^{CE}$ for both PMTs against the relative distance to PMT2 (see text for detailes). 2009 data.
\label{FitNdLS2009}}
\end{figure}

To address the issue of NdLS stability, which is crucial among the others, we have kept the cell filled with NdLS along $\sim$1 year and then performed a new set of measurements scanning as usual the cell along its main axis by the $^{137}$Cs source. In Fig.~\ref{FitNdLS2009} we plot the quantity $L_{1/2}^{CE}$ evaluated in 2009 spectra as a function of the source distance from each PMT; the only operation done on the NdLS prior to the measurements was nitrogen sparging to remove the oxygen that unavoidably entered in the cell, as the cell sealing is not guaranteed along such a long time. It is relevant to add that the laboratory where the cell was stored along the year was not air conditioned, therefore the temperature reached quite high values ($\sim$35$^o$C along the summer months).
The fitted uncorrected, for actual optical cell path and light propagation effects, $\Lambda$ value is found to be 87$\pm$3 cm to be compared with 74$\pm$3 cm fitted from the May 2008 data set; therefore we can state that the scintillator is stable along the observation period.

\begin{figure}[t]
\centerline{
\includegraphics[width=0.6\textwidth,angle=0]{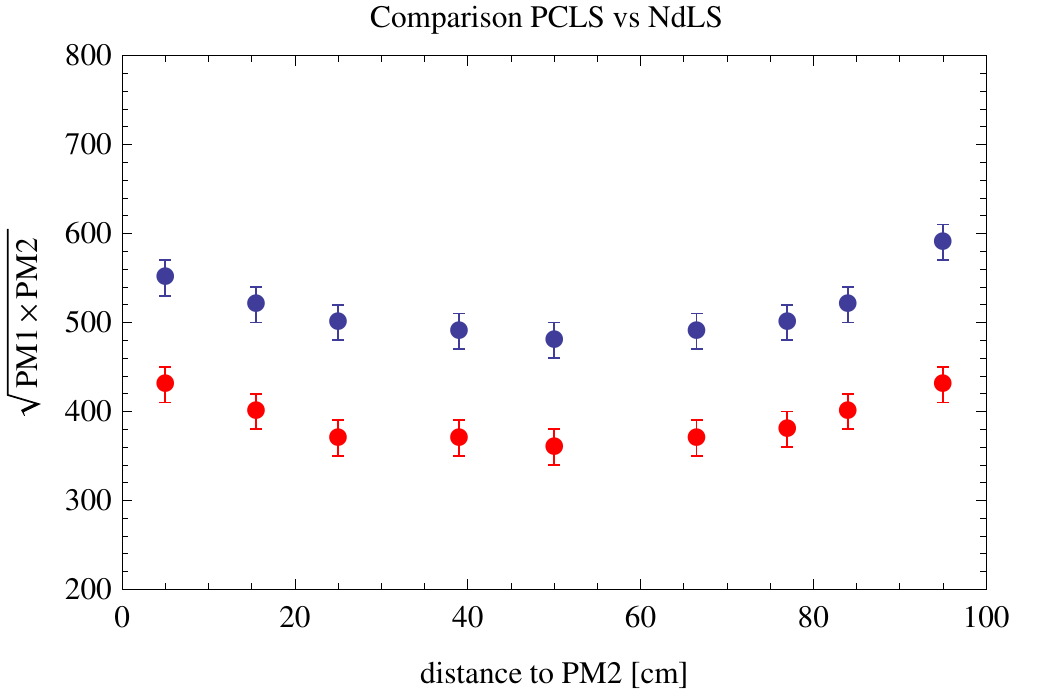}
}
\caption{
Distribution of $L_{1/2}^{CE}$ from the spectra built for $\sqrt{Q_{PM1} \cdot Q_{PMT2}}$ against the relative distance to PMT2 (see text for detailes). 
\label{PCNdLS}}
\end{figure}

To determine the light yield, we compute the quantity: 
 \begin{equation}
  \sqrt{Q_{PMT1} \cdot Q_{PMT2}} \propto LY\cdot e^{-\frac{L}{\Lambda}}
 \label{eqno1}
 \end{equation}
 where $Q_i$ is the charge of an event, $LY$ is the light yield (p.e./MeV), $L$ the length of the cell and $\Lambda$ effective attenuation length for the 1-m cell.
Then we plot it as a function of the source position taking as reference PMT2 as shown in  Fig.~\ref{PCNdLS}.
In particular, we are interested in the ratio:
 \begin{equation}
\frac{\sqrt{Q_{PMT1}^{NdLS} \cdot Q_{PMT2}^{NdLS}}}{\sqrt{Q_{PMT1}^{LS} \cdot Q_{PMT2}^{LS}}} =\frac{LY_{NdLS}}{LY_{LS}}\cdot 
e^{-\frac{L}{2}\cdot\frac{\Lambda_{LS}-\Lambda_{NdLS}}{\Lambda_{LS}\cdot\Lambda{NdLS}}}
 \label{eqno2}
 \end{equation}
The ratio in Eq~(\ref{eqno2}) allows to determine the LY in the NdLS relative to that of the undoped LS, that is by how much the Nd reduces the scintillation light. 
This is shown in Fig.~\ref{PCNdLS}, where it can be seen that the NdLS reduces by $\sim$25\% the light yield of the pure LS. This value is consistent with the value measured in the small vials at the time of production.

\section{Radiopurity and purification methods}
The intrinsic radiopurity of Nd has been measured with HP-Ge detectors at the Gran Sasso underground facility~\cite{HpGe}. On a sample of 50.2~g of commercial Nd$_2$O$_3$ interactions of photons produced by internal radioactive contaminants have been recorded for 37.3~days. In Table~\ref{HpGetab} we report the results of the measurement.

 \begin{table}[ht]
\begin{tabular}{c c c c c c c c }
\hline
$^{228}$Ra & $^{228}$Th  & $^{226}$Ra  & $^{234}$Th  & $^{234}$Pa$^m$ & $^{235}$U & $^{40}$K & $^{137}$Cs  \\
  ppb             &  ppb                &    ppb             &   ppb              &    ppb                       &     ppb         &   mBq/k &  mBq/kg          \\
\hline
6.6$\pm$1.8 & 8.5$\pm$1.1 & 0.39$\pm$0.23 & $<$22 & $<$9.5 & $<$9.1 & $<$86 & 9.7$\pm$2.7\\
\hline
\end{tabular}
\caption{Radiopurity measurement of commercial Nd$_2$O$_3$ used for the NdLS production by means of HP-Ge. Upper limits are at 90\% C.L.}
\label{HpGetab}
\end{table}

The most important source of background we are concerned about for applications of the NdLS is $^{232}$Th. From data in Table~\ref{HpGetab} we conclude that in the commercial  $Nd_2O_3$ a contamination at the level of 8~ppb in $^{232}$Th is present. This turns to about 30~mBq/kg.
A second set of measurements were carried out by ICP-MS on different packages of the same production lot. Samples were diluted in a 10~ml $HNO_3$ solution at 2.5\%. In Table~\ref{ICPMStab} we report results.

 \begin{table}[ht]
 \begin{center}
\begin{tabular}{c c c c }
\hline
Nd sample   & 104 mg  & 237 mg  & 0.067 mg \\
\hline
$^{232}$Th [ppb] & 5$\pm$1  & 7$\pm$2 &  6$\pm$1\\
$^{238}$U [ppb] & $<$ 0.2 & $<$0.3  & $<$0.3  \\
\hline
\end{tabular}
\caption{Radiopurity measurement of commercial Nd$_2$O$_3$ used for the NdLS production from ICP-MS. Upper limits are at 68\% C.L.}
\label{ICPMStab}
\end{center}
\end{table} 
 
Both methods applied to determine the intrinsic Th contamination in Nd$_2$O$_3$ have shown similar results. In order to reduce the Th contamination we have developed and tested two purification methods. In a first case we have used a liquid-liquid purification. In particular, at room temperature we perform a liquid-liquid/ extraction using NdCl$_3$ in Hydrochloric acid solution and an equal volume of TOPO diluted in toluene. Mixing lasted for about 2 minutes. Using 0.2M TOPO in toluene  and starting from a sample spiked with 200 ppb of Th we have reached a final contamination of 2 ppb, that is a purification factor of $\sim 100$. 

As an alterative to the liquid-liquid extraction method a selective ion exchange column procedure using 
UTEVA resin has been attempted. The active group in UTEVA resin is  the dipentyl-pentyl -phosphonate that
selectively extracts thorium and uranium from a solution containing an high (3M) concentration of nitric acid. 
A 15\% Nd(NO$_3$)$_3$ solution in 3M nitric acid artificially contaminated with Th is passed through a column filled with the UTEVA resin that
retains Th and U. Immediately after, the column is washed with 3M nitric acid to recover most part of the
Neodimium. The Neodimium loss is less than 1\% and the purification factor, calculated measuring the Th concentration both in the un-purified and purified solution by mean ICP-MS, is about 200. The column can be used without significant loss of efficiency to purify about 15g of Nd for 
1g of resin. The column can be regenerated for some (less then 10) times using strong HCl (6M) to
elute the thorium and the uranium 
from the column, then washing the column  with high purity water and conditioning again the resin with 3M nitric acid.
The advantages of this procedure are the higher purification factor that can be obtained and the fact that the method is well suited for a large-scale continuous operation. The disadvantages are the high cost of the resin and the fact that the Neodimium is recovered in a strongly acidic 
solution that is not usable for the Nd carboxylate synthesis. For this reason a further step is needed to neutralize the nitric acid, precipitate the Nd as hydroxide with the aid of urea, purify the precipitate from the nitric ion and finally to convert the hydroxide to chloride. This step complicate further the procedure so that only a marginal advantage remains from the significant (two-fold) increase of the purification factor.

\section{Neutrinoless double beta decay with a 12 tons Nd-loaded scintillator}
The neutrinoless double beta decay ($0\nu2\beta$), $(A,Z) \rightarrow (A,Z+2)+2e^-$, will reveal the Majorana nature of neutrinos. For a given candidate nucleus (A,Z) the quantity to be measured is the decay rate: $T^{-1}_i = G_i |M_i|^2m^2_{\beta\beta}$, where $G_i$ is a determinable phase-space factor, $M_i$ the nuclear matrix element (NME) and $m^2_{\beta\beta}$ the so-called effective Majorana neutrino mass. The quantity to be measured contains two unknowns. Therefore, at least two measurements are needed to determine  $m^2_{\beta\beta}$. Extra measurements are needed to test the nuclear model. At present a claim of evidence comes from the search in $^{76}$Ge~\cite{Klapdor}. This claim turns into a 90\% confidence interval for $m^2_{\beta\beta}$ equal to [0.17,0.45]~meV~\cite{Lisi}. Therefore, 100~meV can be taken as a reference value for coming experiments to probe the present finding.

The idea to search for neutrinoless double beta decay with a Nd-loaded liquid scintillator is not new~\cite{SNO+}. In this case a large mass of Nd is used in 1~kton of liquid scintillator and this allows to search for such a rare decay process in spite of the poor energy resolution when comparing with other detectors~\cite{CUORE,GERDA}. In the following we propose a similar measurement but using a much smaller detector. In particular, we aim to merge the Borexino technology~\cite{Borexinodetector}, which has shown the feasibility to purify 1~kton of liquid scintillator at the level of 10$^{-17}$~g/g in $^{238}$U and $^{232}$Th, with the results established in our measurements and considering the performances of new PMTs with high quantum efficiency (QE) and low radioactivity ($<$ 100~mBq)~\cite{newPMTs}. 
Taking into account the results obtained and reported in the previous Sections, we have studied the possibility to search for the Nd zero-$\nu$ mode double beta decay with a 10-ton scale target mass of NdLS based on PC. In particular, we can assume the same bulk radiopurity reached by Borexino because we are using the same solvent and fluor. Therefore, our basic assumptions are a PC-based organic liquid scintillator with 1.5~g/l of PPO and 6.5~g/l of Nd. As described above this NdLS will have a light yield reduced by a 25\% with respect to pure PC-based scintillator and an attenuation length of $\sim$2~m. We make use of 12~tons of NdLS in a configuration similar to the one used in Borexino~\cite{Borexinodetector}. The NdLS is contained in a nylon vessel 1.5~m in radius inside a 4~m radius and 1~cm thick stainless steel sphere filled with distilled PC. As many as possible 3-inch low radioactivity and high quantum efficiency PMTs are installed on the inner surface of the sphere so that a 80\% of coverage is reached. Light concentrators can be used to reduce the total number of PMTs needed to get a large light collection~\cite{Concentrator}. A geometrical amplification factor on the order of 4 or larger can be assumed. Using 4 as many as  2000 PMTs will be needed. The total activity at the inner nylon vessel due to PMTs is estimated to be $\sim$ 2mBq, using 100~mBq/PMT and $\lambda_{PC}$=27.2~cm for the absorption length of 2.6~MeV photons in PC. The stainless steel sphere will produce about 0.4~mBq using a steel contamination of $\sim$2~mBq/kg~\cite{GerdaSteel}. So, a maximum activity of about 2.4~mBq will be compared with the rate of the two-$\nu$ decay mode. For such a detector we can determine the light yield, $LY$, as: $LY = G_\lambda \times G_c \times QE \times \eta \times Q_{Nd} \times N_\gamma$, where $G_\lambda$=0.62 accounts for the reduction of light collection efficiency due to the finite attenuation lenght, $G_c$=0.80 is the geometrical light collection, $QE$=0.40\footnote{The quantum efficiency depends on the wavelength. Using data from~\cite{newPMTs} we have determined that the average value of QE is found at 395nm. However, the number of photoelectrons registered depend upon the QE, the emission spectrum and the trasmittance. As it is shown in Fig.~\ref{assorbanza e emissione} the NdLS has low attenuation between 370nm and 420nm, when this is taken into account the average sits at 391nm, where QE$\sim$0.40. Therefore, our assumption is well justified.}, $\eta$=0.90 the PMT collection efficiency on the first dynode and $N_\gamma \sim$10$^4$ photons/MeV for a PC-based liquid scintillator. With the above assumptions we obtain at a first approximation about 1340~p.e./MeV and a resolution of 3.5\% FWHM at $Q_{\beta\beta}$=3.367~MeV. For Nd zero-$\nu$ decay mode the half-life, $T^{0\nu}_{1/2}$, in years can be written~\cite{Avignone}: $\frac{1}{T^{0\nu}_{1/2}} =\frac{57\cdot10^{-19}}{m_e^2} m^2_\nu$, where $m_\nu$ is in units of meV and the electron mass, $m_e$, in units of eV. The quantity $T^{0\nu}_{1/2}$ can be used to determine the expected rate for the zero-$\nu$ decay mode. For the present study we do not consider uncertainties on predictions for the transition matrix related to the decay because this is beyond the purpose of this paper.

In our target mass of 12~tons an intrinsic bulk contamination of Th at the level of 10$^{-5}$ppt will produce about 6~cpy from $^{208}$Tl ($Q_\beta$=5~MeV). Only 0.1~cpy will fall within 1$\sigma$ of  $Q_{\beta\beta}$. At a concentration of 6.5~g/l about 92~kg of Nd are needed. For an effective Majorana neutrino mass of 300(100)~meV we expect $\sim$28(3)~cpy due to the zero-$\nu$ decay mode. The two-$\nu$ decay mode gives about $\sim$50~mBq using $T^{2\nu}_{1/2}=9.2\times10^{18}$~years. Therefore, this decay mode will give by far the largest counting rate below 3~MeV. Another important source of background are $^8$B solar neutrinos. These neutrinos will produce $\sim$22 cpy and only about 0.3 are expected within 1$\sigma$ of  $Q_{\beta\beta}$. In Table~\ref{tabsummary} we summarize the expected rates for signal and backgrounds for $m_\nu$=300(100)~meV. In these two cases natural abundance (5.6\%) is considered. For the purpose of our study we have assumed a Th contamination of 1(0.1)~ppt in Nd and considered a 90\% reduction due to $^{212}$Bi-$^{212}$Po tagging and statistical subtraction. In this scenario the Th contamination in the Nd will give about 10(1)~cpd within 1$\sigma$ of  $Q_{\beta\beta}$.

 \begin{table}[ht]
\begin{tabular}{c c c c c c c c c c}
\hline
M$_{LS}$ & M$_{Nd}$  & M$_{^{150} Nd}$  &  FWHM  & $m_\nu$ & $S^{0\nu}$ & $S^{2\nu}$ & $^8$B $\nu_\odot$ & $^{208}$Tl    & $S/\sqrt{B}$  \\
  tons         &  kg       & kg                        &   \%     &  meV      &       cpy      &  cpy            &   cpy                         & cpy       &    \\
\hline
12.3 & 92 & 5.1 & 3.5 & 300 & 19.2 & 0.1 & 0.3 & 10.6 & 5.8\\
12.3 & 92 & 5.1 & 3.5 & 100 & 2.1 & 0.1 & 0.3 & 1.1 & 1.8\\
\hline
\end{tabular}
\caption{The case of a 12 tons detector loaded with 6.5 g/l of Nd at natural abundance. All rates are given within 1$\sigma$ of  $Q_{\beta\beta}$.}
\label{tabsummary}
\end{table}

The case reported in Table~\ref{tabsummary} allows to measure the zero-$\nu$ decay mode at the level of 17\%(40\%) in 3(5) years for $m_\nu$=300(100)~meV.  

\begin{figure}[t]
\centerline{
\includegraphics[width=0.6\textwidth,angle=0]{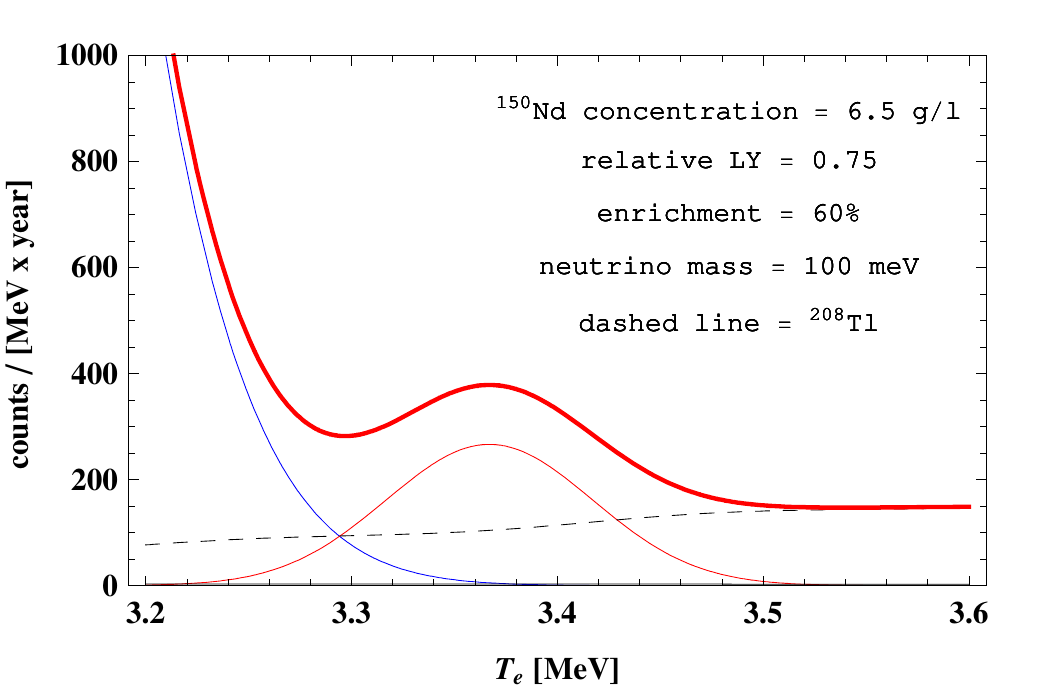}
}
\caption{Expected spectrum for the case of an NdLS enriched at 60\% for $m_\nu$=100 meV.
\label{FigDBD}}
\end{figure}

Much better results can be obtained with Nd enriched in $^{150}$Nd. An enrichment factor of 60\% would give $S^{0\nu}$=23~cpy with 55~kg of $^{150}$Nd within 1$\sigma$ and for $m_\nu$=100~meV. Such a possibility could be within experimental and technological reach due to the limited $^{150}$Nd mass needed~\cite{Labozin}. Moreover, for a small target mass the background due to solar neutrinos is  negligible and not as important as it could be in the case of a larger detector such as SNO+.
In conclusion with a target mass of 12~tons of LS, a 60\% Nd enrichment and 1ppt of Th in Nd a measurement at the level of 15\%(30\%) seems feasible in 3(5) years of data taking for $m_\nu$=100(50) meV. In Table~\ref{tabsummaryDBD} and in Fig.~\ref{FigDBD} we summarize our results for the case of the enriched detector.

 \begin{table}[ht]
\begin{tabular}{c c c c c c c c c c}
\hline
M$_{LS}$ & M$_{Nd}$  & M$_{^{150} Nd}$  &  FWHM  & $m_\nu$ & $S^{0\nu}$ & $S^{2\nu}$ & $^8$B $\nu_\odot$ & $^{208}$Tl    & $S/\sqrt{B}$  \\
  tons         &  kg       & kg                        &   \%     &  meV      &       cpy      &  cpy            &   cpy                         & cpy       &    \\
\hline
12.3 & 92 & 55 & 3.5 & 100 & 23 & 1.0 & 0.3 & 10.6 & 6.6\\
12.3 & 92 & 55 & 3.5 & 50 & 5.7 & 1.0 & 0.3 & 10.6 & 1.7\\
\hline
\end{tabular}
\caption{The case of a 12 tons detector loaded with 6.5 g/l of Nd at 60\% enrichment and 1ppt Th in Nd. All rates are given within 1$\sigma$ of  $Q_{\beta\beta}$.}
\label{tabsummaryDBD}
\end{table}

\section{Conclusions}
Neutrinoless double beta decay is one of the most important research goals in Astroparticle Physics at present. In this paper we have shown how this search can be carried out by means of a highly doped liquid organic scintillator with a target mass of about 10 tons and using $^{150}$Nd. We have reported results of measurements carried out to determine the optical properties, time stability and photon yield for a 6.5 g/l Nd doped liquid scintillator based on pseudocumene. In particular, with our experimental results we have shown that a target mass of about 12 tons with about 60 kg of enriched Nd could reach a sensitivity in the range of 50-100~meV in the effective neutrino mass. We have also reported about purification methods studied to reduce the intrinsic $^{232}$Th contamination in commercial Nd. We believe our results could pave the way for further studies with the aim to better characterize the feasibility of the technique presented for neutrinoless double beta decay search. 

\section{Acknowledgments}
We thank Frank Calaprice and Daniele Montanino for useful discussions about neutrinoless double beta decay search based on liquid scintillators.
We thank Matthias Laubenstein responsible of the low radioactivity laboratory at Gran Sasso for his effort made on Nd samples. This work is supported by INFN (Istituto Nazionale di Fisica Nucleare) through the MetaLS group. 




\begin{thebibliography}{00}
\bibitem{MetaLS} I. Barabanov et al., J. Phys.: Conf. Ser. 136 042088 (2008).
\bibitem{YbIGdLS} N.A. Danilov et al., Radiochemistry 45 (2003) 140--146.
\bibitem{InLS1} N.A. Danilov et al., Radiochemistry 47 (2005) 487--493.
\bibitem{InLS2} C. Buck et al., Nucl. Phys. B (Proc. Suppl.) 143 (2005), 487 
\bibitem{GdLS1} N.A. Danilov et al., Radiochemistry 51-3 (2009) 274-283.
\bibitem{GdLS2} A. di Vacri,  Progress in Particle and Nuclear Physics 57 (1), pp. 296.
\bibitem{GdLS3} I.R. Barabanov et al. http://arxiv.org/abs/0803.1577v1 
 \bibitem{BorexinoNIM440} A. Alimonti et al., Borexino coll., NIM A 440 (2000) 360-371.
 \bibitem{HpGe}C. Arpesella et al., Borexino coll., Astrop. Phys.18:1-25,2002. 
  \bibitem{Klapdor} H.V. Klapdor-Kleingrothaus and L.V. Krivisheina, Mod. Phys. lett. A21: 1547-1566, 2006.
  \bibitem{Lisi} A. Faessler et al., J.Phys.G35:075104,2008.
 \bibitem{SNO+} M.C. Chen, SNO+ coll., 34$^{th}$ Conference on High Energy Physics, Philadelphia, 2008.
 \bibitem{CUORE} C. Arnoboldi et al., CUORE coll., NIM A 518 (2004) 775-798.
 \bibitem{GERDA} S. Schoenet et al., GERDA coll., Phys.Atom.Nucl.69:2101-2108,2006.
 \bibitem{Borexinodetector} G. Alimonti et al., Borexino coll., NIM A 600 (2009) 568-593.
 \bibitem{newPMTs} Hamamatsu PMTs, Super and Ultra Bialkali photomultiplier tube series.
 \bibitem{Concentrator} L. Oberauer et al., NIM A 530 (2004) 453-462.
 \bibitem{GerdaSteel} W. Maneschg et al., NIM A 593:448-453,2008.
 \bibitem{Avignone} F.T. Avignone, Nucl. Phys. b (Proc. Suppl.) 143 (2005) 233-239.
 \bibitem{Labozin} A.P. Babichev et al., Quantum Electronics 35(10) 879-890 (2005).


\end{thebibliography}
\end{document}